\documentclass{osa-article}

\journal{osajournal}
\articletype{Research Article}
\begin{document}

\title{Photon statistics of superbunching pseudothermal light}

\author{Chaoqi Wei,\authormark{1} Jianbin Liu,\authormark{1, 2,*}  Xuexing Zhang, \authormark{1} Rui Zhuang, \authormark{1} Yu Zhou, \authormark{3} Hui Chen, \authormark{1} Yuchen He, \authormark{1} Huaibin Zheng, \authormark{1} and Zhuo Xu \authormark{1}}

\address{\authormark{1}Key Laboratory of Multifunctional Materials and Structures, Ministry of Education \& International Center for Dielectric Research, School of Electronic Science and Engineering, Xi’an Jiaotong University, Xi'an 710049, China\\
\authormark{2}The Key Laboratory of Weak Light Nonlinear Photonics (Nankai University, Tianjin 300457), Ministry of Education, China\\
\authormark{3}MOE Key Laboratory for Nonequilibrium Synthesis and Modulation of Condensed Matter, Department of Applied Physics, Xi’an Jiaotong University, Xi’an, Shaanxi 710049, China}

\email{\authormark{*}liujianbin@xjtu.edu.cn} %% email address is required



\begin{abstract}
Superbunching pseudothermal light has important applications in studying the second- and higher-order interference of light in quantum optics. Unlike the photon statistics of thermal or pseudothermal light is well understood, the photon statistics of superbunching pseudothermal light has not been studied yet. In this paper, we will employ single-photon detectors to measure the photon statistics of superbunching pseudothermal light and calculate the degree of second-order coherence. It is found that the larger the value of the degree of second-order coherence of superbunching pseudothermal light is, the more the measured photon distribution deviates from the one of thermal or pseudothermal light in the tail part. The results are helpful to understand the physics of two-photon superbunching with classical light. It is suggested that superbunching pseudothermal light can be employed to generate non-Rayleigh temporal speckles.
\end{abstract}

%%%%%%%%%%%%%%%%%%%%%%%%%%  body  %%%%%%%%%%%%%%%%%%%%%%%%%%
\section{Introduction}\label{introduction}
Thermal light plays an essential role in the development of modern quantum optics \cite{glauber-rmp}. For instance, the first experimental observation of the second-order interference of light was conducted with thermal light by Hanbury Brown and Twiss in 1956 \cite{hbt,hbt-1}, which was latter known as two-photon bunching of thermal light. However, implementing the second- and higher-order interference of light with true thermal light is extremely difficult \cite{hbt-book}, due to the coherence time of true thermal light is much shorter than the response time of the fastest photo-detector available, let alone the low degeneracy parameter of true thermal light requires long collecting time in experiments \cite{mandel-book}. In order to solve  this problem, Martienssen and Spiller invented pseudothermal light to simulate the properties of true thermal light by inputing single-mode continuous-wave laser light onto a rotating groundglass \cite{martienssen}.  Pseudothermal light simplifies the experimental requirements a lot compared to the ones with true thermal light due to the coherence time and degeneracy parameter of pseudothermal light can be easily manipulated \cite{martienssen}. Since its invention, pseudothermal light has been extensively applied in the second- and higher-order interference experiments such as ghost imaging \cite{gi-2004,gi-2005,gi-2009,gi-2015}, two-photon interference \cite{ti-2005,ti-2006},  intensity interferometer \cite{ii-2010,ii-2018}, quantum erasers \cite{dc-2014}, \textit{etc}.

Recently, we modified the well-known pseudothermal light source into superbunching pseudothermal light source by adding more than one rotating groundglass \cite{zhou-2017} or modulating the intensity of laser light before the rotating groundglass \cite{zhou-2019,arxiv-2021}. The degree of second-order coherence of superbunching pseudothermal light can be much larger than the one of thermal or pseudothermal light \cite{zhou-2017, zhou-2019}, indicating that the photon distributions of these two types of light are different. The photon statistics of thermal or pseudothermal light follows geometrical distribution \cite{loudon-book}, which is the feature distribution of Rayleigh speckles \cite{goodman-speckle}. Rayleigh speckles are the result of superposition of large number of independent fields with random amplitudes and uniform random phases between 0 and 2$\pi$. Non-Rayleigh speckles can be generated when the phases of the superposed fields are correlated \cite{cao-2014, cao-2017, cao-2018, zhang-2019} or the number of superposed fields is not large enough \cite{rouge-2017}. However, most experiments generate non-Rayleigh speckles in spatial domain \cite{cao-2014, cao-2017, cao-2018, zhang-2019,rouge-2017}.  There are only a few researches studying non-Rayleigh speckles in temporal domain. Straka \textit{et. al.} employed acousto-optical modulator to generate arbitrary classical photon statistics, in which the intensity across the whole light beam does not change  \cite{temporal-2018}. In our early experiments, we find a way to generate non-Rayleigh speckles in temporal domain by superposing a large number of field and varying the intensity of the input laser light in time \cite{zhou-2017, zhou-2019}. The generated non-Rayleigh temporal speckles is different from the one in Ref. \cite{temporal-2018} since the intensities across the light beam are not constant. The superbunching pseudothermal light can be employed to study multi-photon interference \cite{ou-book} and high-visibility temporal ghost imaging \cite{tgi-2018}. In our early researches, we mainly focused on studying the degree of second- and higher-order coherence of superbunching pseudothermal light \cite{zhou-2017, zhou-2019, arxiv-2021, zhou-josab}. The study on the photon statistics of superbunching pseudothermal light is missing, which contains more information about the properties of light than the second- and higher-order degree of coherence. In this paper, we will study the photon statistics of superbunching pseudothermal light in detail, which is helpful to understand the properties of this newly proposed light.

The remaining parts of this paper are organized as follows. Section \ref{theory} contains the theory of how to calculate the degree of second-order coherence from photon statistics. The experiments measuring the photon statistics of superbunching pseudothermal light are in Sect. \ref{experiments}. The discussions and conclusions are in Sects. \ref{discussions} and \ref{conclusions}, respectively.

\section{Theory}\label{theory}
The degree of second-order coherence, $g^{(2)}(0)$, can be employed to distinguish different types of light \cite{loudon-book}. For coherent light, $g^{(2)}(0)$ equals 1, which indicates that the detections of two photons are independent. For thermal light,  $g^{(2)}(0)$ equals 2, which means that the detections of two photons are correlated. $g^{(2)}(0)$ can not be less than 1 in classical theory. Light with $g^{(2)}(0)$ being less than 1 is non-classical light. For instance, $g^{(2)}(0)$ equals 0 for single-photon state, which is absolutely impossible in classical theory \cite{loudon-book}. This is the reason why knowing the value of the degree of second-order coherence of light is enough in most cases. However, things become different in some cases. For instance, the degree of second-order coherence of entangled photon pair light can be much larger than 2 \cite{klyshko-1970, burnham-1970}, which belongs to non-classical light based on Glauber-Sudarshan criterion \cite{glauber,glauber1, sudarshan}.  On the other hand, $g^{(2)}(0)$ can also be much larger than 2 for classical light \cite{zhou-2019,temporal-2018}. It is not able to distinguish these two types of light if only the value of the degree of second-order coherence is given. More information, such as photon statistics, is needed to distinguish different types of light once the values of $g^{(2)}(0)$ are at the same level.

Photon distribution, also known as photon number distribution, contains more information than the degree of second-order coherence of light, which can be employed to calculated $g^{(2)}(0)$ of  light . Photon distribution is usually measured with single-photon detectors. The degree of second-order coherence of light can be calculated via photon distribution \cite{loudon-book}
\begin{equation}\label{g2}
g^{(2)}(0)=\frac{\langle n(n-1)\rangle}{\langle n\rangle^2},
\end{equation}  
in which $n$ is the number of photons and $\langle \rangle$ represents ensemble average. In experiments, the photon number $n$ can be replaced by the measured number of photons,
\begin{equation}\label{mtT}
\hat{M}(t,T)=\int_t^{t+T}dt'\hat{a}^{\dag}(t')\hat{a}(t'),
\end{equation}  
where $\hat{M}(t,T)$ is the number of the measured photons between time $t$ and $t+T$, $T$ is the width of collecting time window, $\hat{a}^{\dag}(t')$ and $\hat{a}(t')$ are the photon creation and annihilation operators, respectively. Substituting Eq. (\ref{mtT}) into Eq. (\ref{g2}), the calculated degree of second-order coherence of light can be expressed via the measured photon counts \cite{loudon-book}
\begin{equation}\label{g2m}
g_c^{(2)}(0)=\frac{\langle \hat{M}(t,T)(\hat{M}(t,T)-1)\rangle}{\langle \hat{M}(t,T)\rangle^2},
\end{equation}
in which the value of $g_c^{(2)}(0)$ will coincide with the one of $g^{(2)}(0)$ when $T$ is much shorter than the coherence time and the collecting area of the detector is much smaller than the transverse coherence area of the measured light \cite{mandel-po}. Otherwise, the calculated $g_c^{(2)}(0)$ may be different from $g^{(2)}(0)$, since different modes of the light has to be taken into account. For instance, $g_c^{(2)}(0)$ approximately equals 1 when $T$ is much larger than the coherence time of light, no matter what type of light was measured. In fact, this is the reason why it is difficult to do the second- and higher-order interference experiments with true thermal light, since the coherence time of true thermal light is usually much shorter than the response time of the fastest detector available \cite{hbt-book}.

\section{Experiments}\label{experiments}

\begin{figure}[htb]
\centering
\includegraphics[width=80mm]{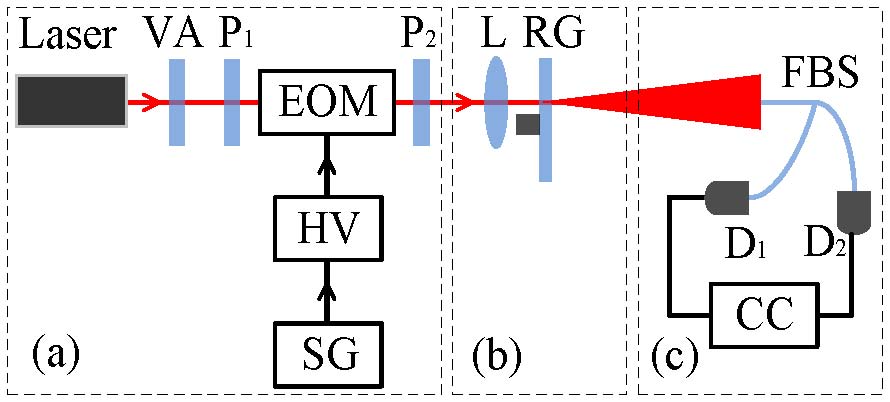}
\caption{(Color online) Experimental setup to measure photon statistics of superbunching pseudothermal light. Laser: single-mode continuous-wave laser. VA: variable attenuator. P: polarizer. EOM: electro-optical modulator. HV: high-voltage amplifier. SG: signal generator. L: lens. RG: rotating groundglass. FBS: 1:1 single-mode fiber beam splitter.  D$_1$ and D$_2$: single-photon detectors. CC: two-photon coincidence count detection system. Please see text for detail descriptions.}\label{1}
\end{figure}

The experimental setup to measure photon distribution of superbunching pseudothermal light is shown in Fig. \ref{1}, which is similar as the one in our earlier research \cite{zhou-2019}. The experimental scheme can be divided into three parts. Figure \ref{1}(a) provides intensity modulated laser light beam by employing an electro-optical modulator (EOM) in combinations with two orthogonally polarized polarizers (P$_1$ and P${_2}$), a high-voltage amplifier (HV), and a signal generator (SG). The employed laser is single-mode continuous-wave laser with central wavelength at 780 nm and frequency bandwidth of 200 kHz. Figure \ref{1}(b) is a typical scheme for pseudothermal light source consisting of a focus lens (L) and a  rotating groundglass (RG) \cite{martienssen}. The lens is employed to control the spot size of laser light on RG, so that the transverse coherence area of the scattered light can be varied \cite{martienssen}. Figure \ref{1}(c) is a Hanbury Brown-Twiss (HBT) interferometer consisting of a $1:1$ fiber beam splitter (FBS), two single-photon detectors (D$_1$ and D$_2$), and a two-photon coincidence counting system (CC). Photon distribution of superbunching pseudothermal light is measured when one single-photon detector is on. The reason why a HBT interferometer is employed is that we can measure $g^{(2)}(0)$ directly and compare it with the calculated value via the measured photon distribution.

\begin{figure}[htb]
\centering
\includegraphics[width=100mm]{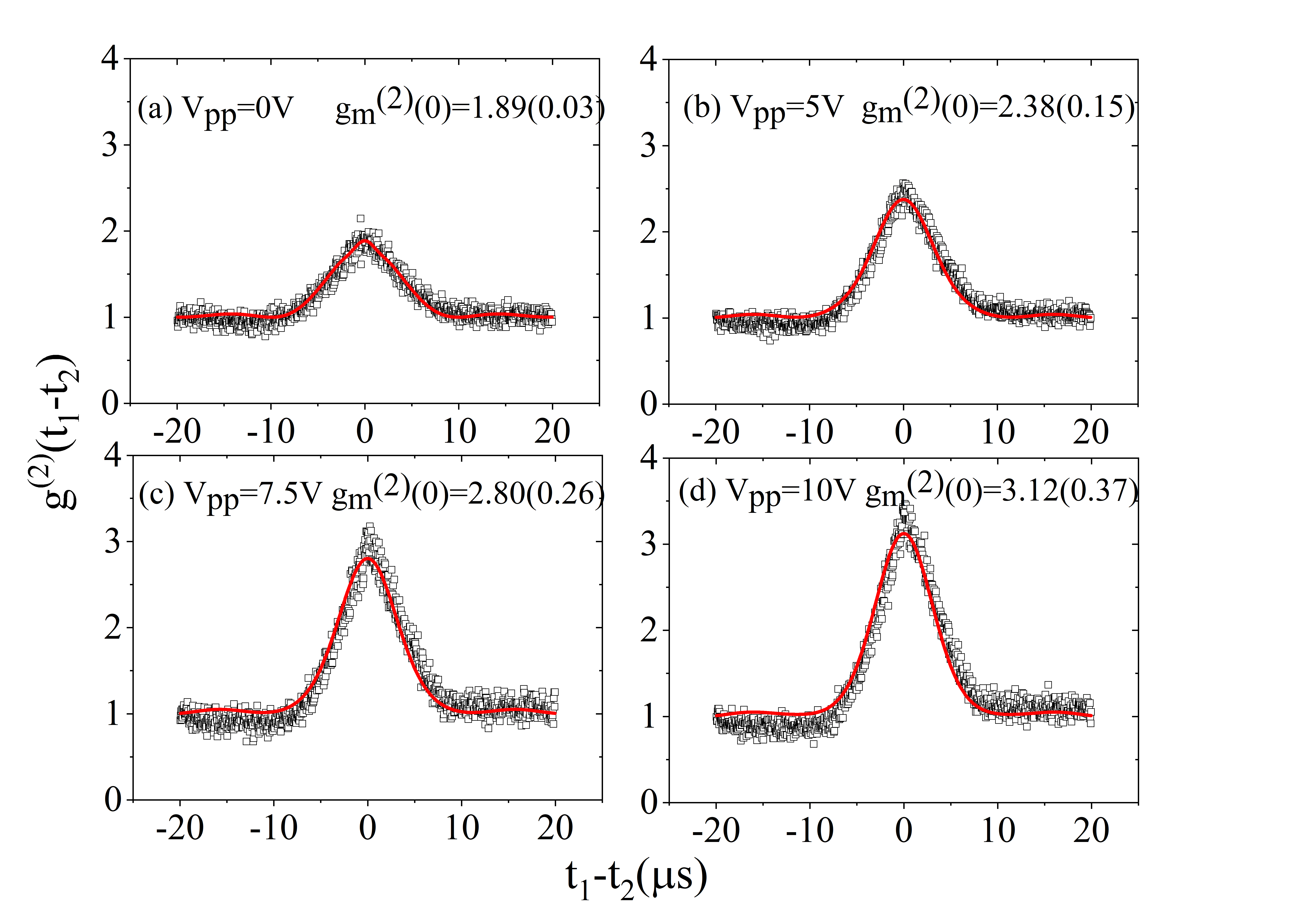}
\caption{Second-order temporal coherence functions of pseudothermal (a) and superbunching pseudothermal (b - d) light. The EOM is driven by the amplified white noise signal generated by SG. The empty squares are experimental results and the red curves are theoretical fittings of the data. V$_\text{pp}$ is the peak-to-peak voltage of the generated white noise. $g_m^{(2)}(0)$ is the measured degree of second-order coherence. (a) - (d)  show the normalized second-order coherence functions when the peak-to-peak voltage of the generated white noise equals 0, 5, 7.5, and 10 V, respectively.}\label{2}
\end{figure}

Figure \ref{2} shows the measured the second-order temporal coherence functions of pseudothermal and superbunching pseudothermal light by applying white noise generated by SG to modulate the intensity of the incident laser light before RG.  $g^{(2)}(t_1-t_2)$ is the normalized second-order temporal coherence function via HBT interferometer. $t_1-t_2$ is the time difference between two single-photon detection events within a two-photon coincidence count. $g_m^{(2)}(0)$ is the measured degree of second-order coherence via HBT interferometer. V$_\text{pp}$ is the peak-to-peak voltage of the generated white noise. As the value of V$_\text{pp}$ increases from 0 V to 10 V, $g_m^{(2)}(0)$ increases from $1.89\pm0.03$ to $3.12\pm0.37$, which means that two-photon superbunching is observed in these three different conditions shown in Figs. \ref{2}(b) - \ref{2}(d). The second-order temporal coherence function of pseudothermal light is measured in Fig. \ref{2}(a), which is observed when V$_\text{pp}$ equals 0 V. The reason why $g_m^{(2)}(0)$ equals 1.89 instead of 2 mainly due to some un-scattered laser photons also contribute to the detected two-photon coincidence counts.  When V$_\text{pp}$ is larger than 0, the measured second-order temporal coherence function is a product of two correlation functions with different correlation time. One correlation time is determined by the modulation frequency of EOM. The other correlation time is determined by the rotating groundglass. These two correlation time are calculated to be 1.28 and  4.63 $\mu$s based on the results in Fig. \ref{2}. 

\begin{figure}[htb]
\centering
\includegraphics[width=100mm]{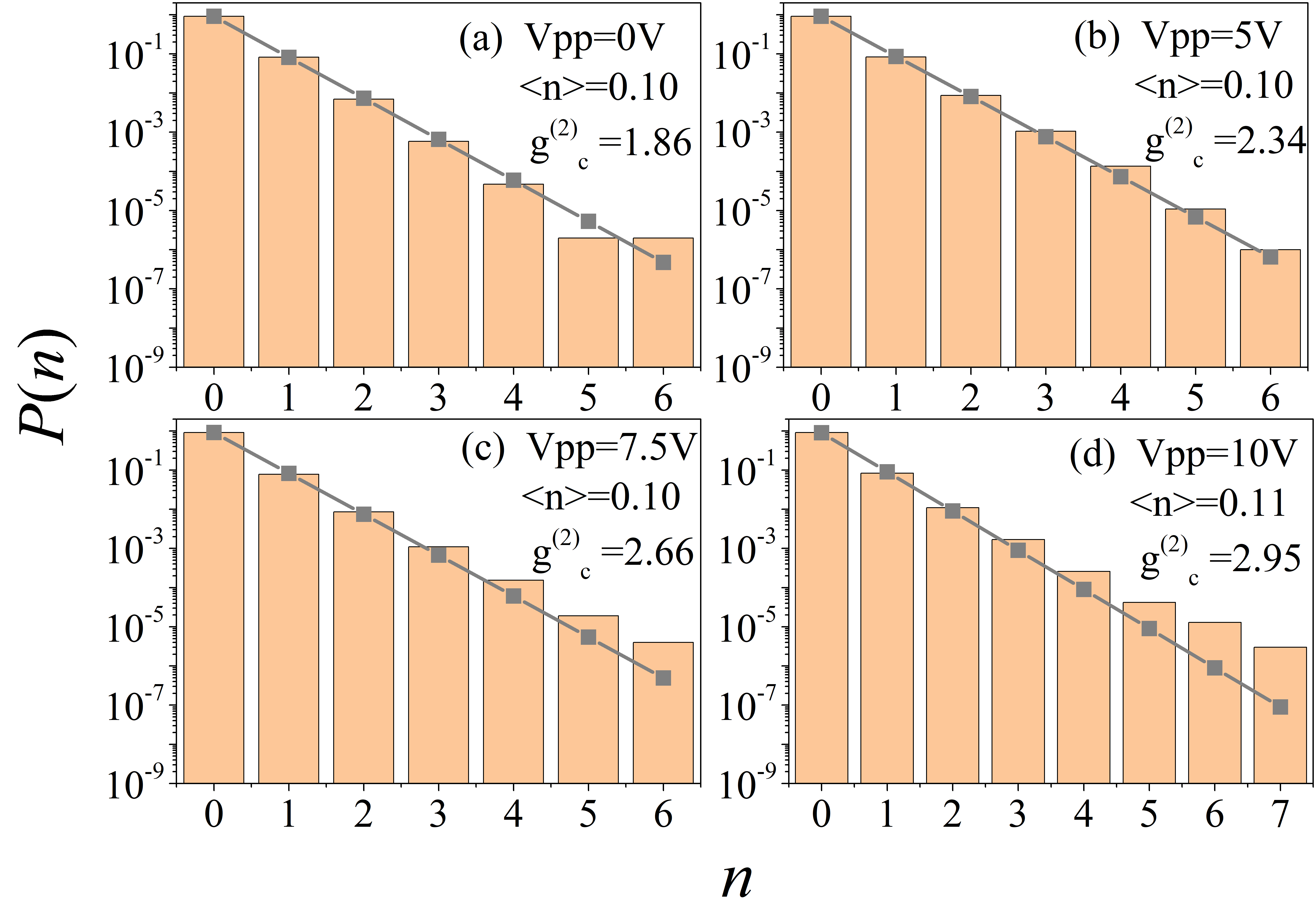}
\caption{Measured photon distribution of pseudothermal (a) and superbunching pseudothermal (b - d) light. (a) corresponds to the photon distribution of pseudothermal light when V$_\text{pp}$ equals 0. (b) - (d) correspond to the photon distribution of superbunching pseudothermal light when V$_\text{pp}$ equals 5, 7.5, and 10V, respectively. $\langle n \rangle$ is the average number of photons in a detection time window.}\label{3}
\end{figure}

Then we measured the photon distribution of pseudothermal light and superbunching pseudothermal in different conditions, which is measured by employing different driving voltage signal generated from the signal generator in the scheme shown in Fig. \ref{1}. V$_\text{pp}$ equaling 0 V is obtained by turning off the signal generator and high voltage amplifier. In this condition, the light source in Fig. \ref{1} is equivalent to a typical pseudothermal light source \cite{martienssen}. Superbunching pseudothermal light source is obtained when V$_\text{pp}$ is larger than a certain value. The diameter of the single-mode fiber is 5 $\mu$m, which is much less than the coherence length of superbunching pseudothermal light in the detection plane. The time window for the photon distribution in Fig. \ref{3} is 5 $\mu$s, which is at the same level of the coherence time of superbunching pseudothermal light. 100k collecting time windows are collected in each measurement. $P(n)$ is the probability of detecting $n$ photons within one collecting time window. $\langle n \rangle$ is the average number of the detected photons in the time window and $g_c^{(2)}(0)$ is the calculated degree of second-order coherence with the help of $P(n)$ and Eq. (\ref{g2m}),
\begin{equation}\label{g2c}
g_c^{(2)}(0)=\frac{\sum_{n=0}^{\infty}{P(n)\times n\times (n-1)}}{[\sum_{n=0}^{\infty}{P(n)\times n}]^2}.
\end{equation}
The lines in Figs. \ref{3}(a) - \ref{3}(d) are geometric photon distribution with the same average photon numbers as the ones in these sub-figures, respectively. The reason why geometric photon distribution is employed to compare with the measured results is that it describes the photon distribution of Rayleigh speckles \cite{goodman-speckle}. As the value of $g_c^{(2)}(0)$ increases, the measured photon distribution of superbunching pseudothermal light is more distributed in the larger number of detected photons part. For instance, the average number of photons both equals 0.10 in Figs. \ref{3}(b) and \ref{3}(c). The calculated degree of second-order coherence, $g_c^{(2)}(0)$, in Fig. \ref{3}(c) is larger than the one in Fig. \ref{3}(b). The probabilities of detecting  5 and 6 photons in Fig. \ref{3}(c) are larger than the ones in Fig. \ref{3}(b). The reason is easy to follow. Larger probabilities to detecting large number of photons means larger photon number fluctuations, which of course will give a larger value of $g_c^{(2)}(0)$ \cite{shih-book}.

\begin{figure}[htb]
\centering
\includegraphics[width=100mm]{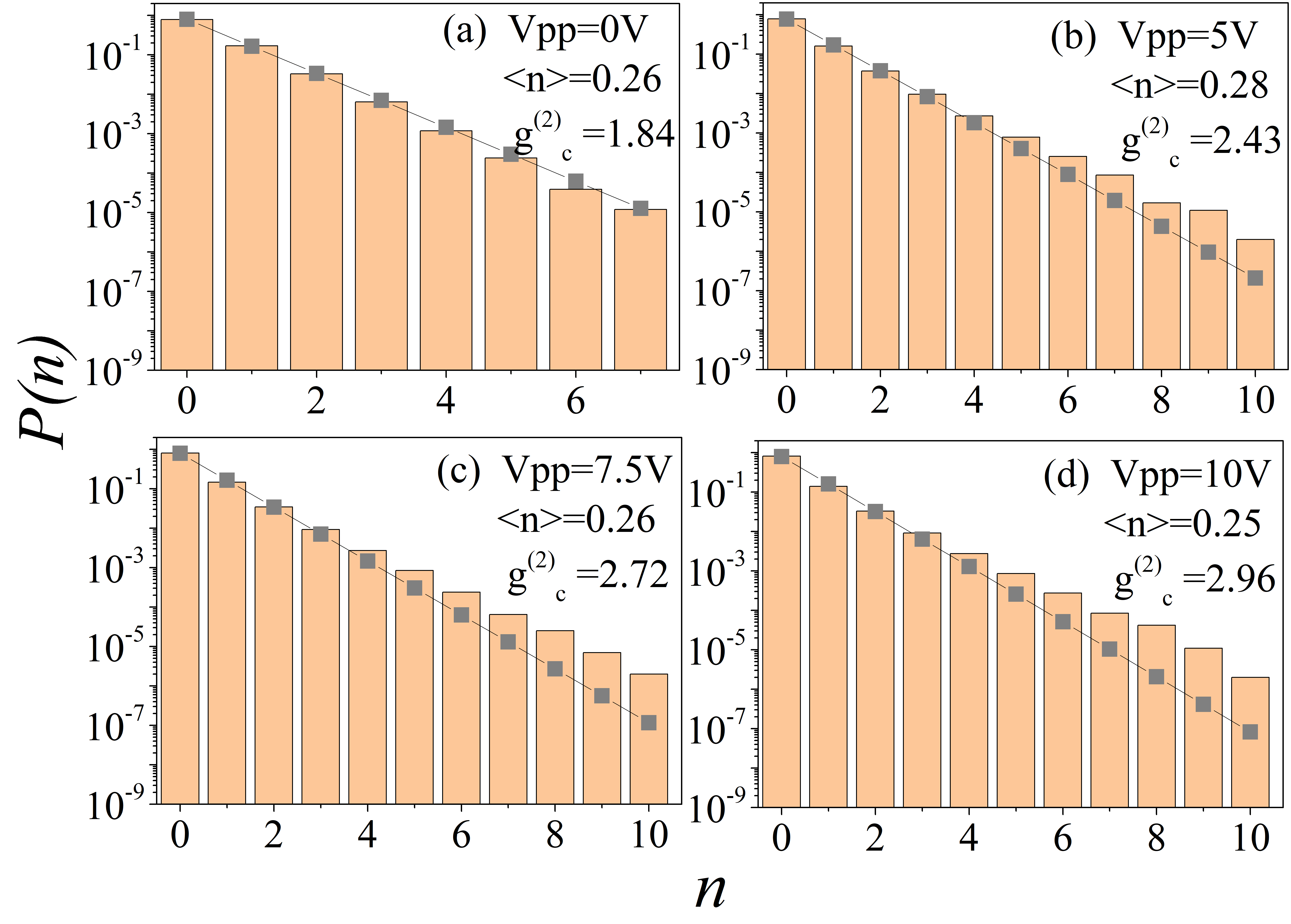}
\caption{Photon distribution of pseudothermal (a) and superbunching pseudothermal (b - d) light when the average number of photons equals about 0.25. The meanings of the symbols are the same as the ones in Fig. \ref{2}.}\label{4}
\end{figure}

\begin{figure}[htb]
\centering
\includegraphics[width=100mm]{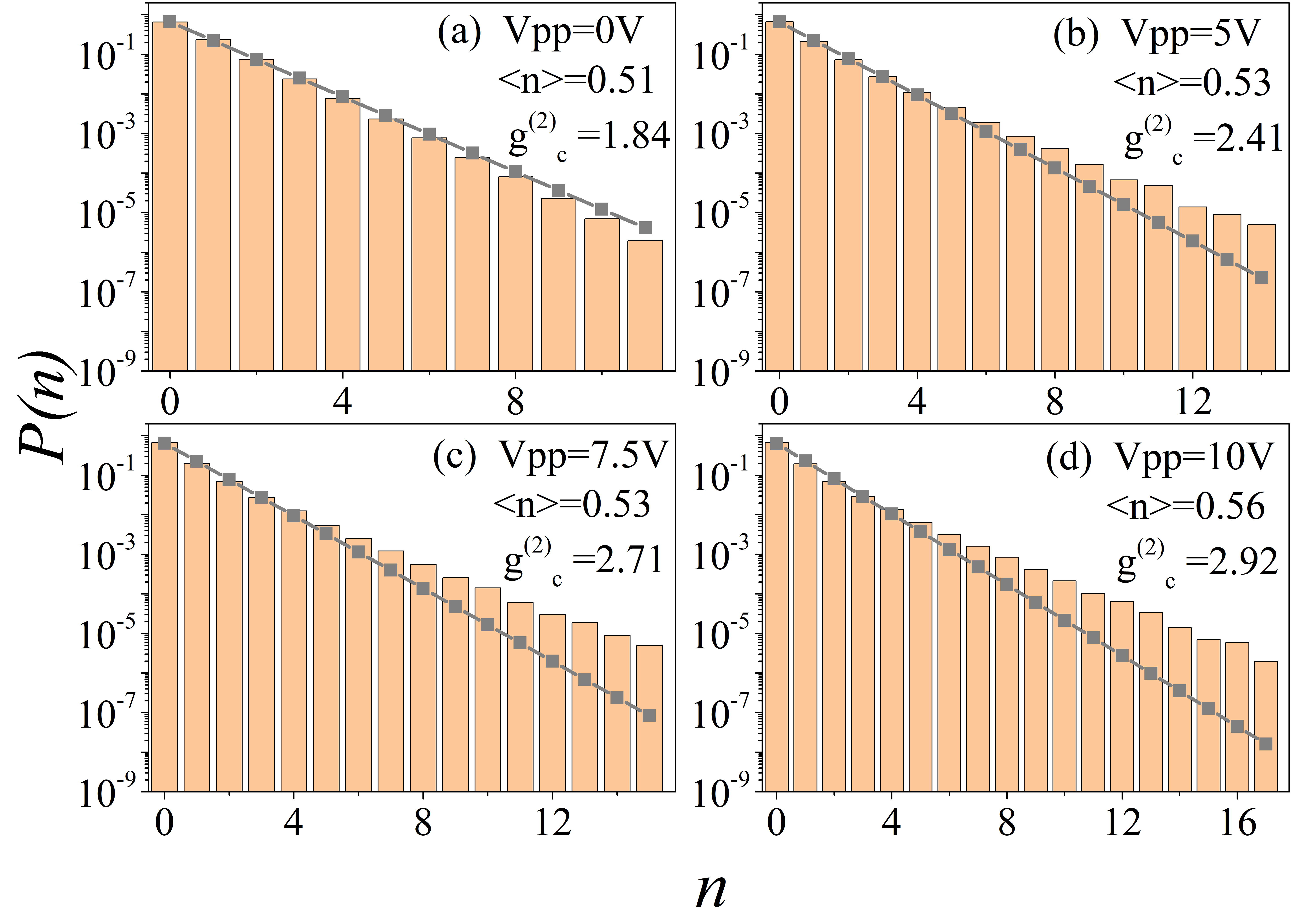}
\caption{Photon distribution of pseudothermal (a) and superbunching pseudothermal (b - d) light when the average number of photons equals about 0.50. The meanings of the symbols are the same as the ones in Fig. \ref{2}.}\label{5}
\end{figure}

Figures \ref{4} and \ref{5} show similar measured photon distributions of pseudothermal and superbunching pseudothermal light as the ones in Fig. \ref{3} except the average number of detected photons are different. The means of all the symbols in Figs. \ref{4} and \ref{5} are the same as the ones in Fig. \ref{3}. The lines are also theoretical results of geometric photon distribution with the same average number in each case. It is concluded that non-Rayleigh photon statistics is observed with superbunching pseudothermal light, which may be treated as another indication of two-photon superbunching besides the degree of second-order coherence exceeding 2.

\section{Discussions}\label{discussions}

In order to compare the results in Figs. \ref{2} - \ref{5}, we summarized the measured and calculated degree of second-order coherence in Table \ref{t1}. The measured degree of second-order coherence in column 2 is obtained in the HBT interferometer, which is showing in Fig. \ref{2} and can be treated as the standard value. The calculated degree of second-order coherence in columns 3 - 5 is obtained from the measured photon distribution. The measured and calculated degree of second-order coherence are consistent within the experimental uncertainty. 

\begin{table}[htb]
\centering
\begin{tabular}[c]{|c|c|c|c|c|}
\hline
$$&$g_m^{(2)}(0)$(HBT)&$g_c^{(2)}(0)$(0.1)&$g_c^{(2)}(0)$(0.25)&$g_c^{(2)}(0)$(0.5)\\
\hline
V$_\text{pp}$=0 V&$1.89\pm0.03$&1.86&1.84&1.84\\
\hline
V$_\text{pp}$=5 V&$2.38\pm0.15$&2.34&2.43&2.41\\
\hline
V$_\text{pp}$=7.5 V&$2.80\pm0.26$&2.66&2.71&2.71\\
\hline
V$_\text{pp}$=10 V&$3.12\pm0.37$&2.95&2.96&2.92\\
\hline
\end{tabular}
\caption{The measured and calculated degree of second-order coherence in different conditions. $g_m^{(2)}(0)$(HBT) is the measured degree of second-order coherence via the HBT interferometer. $g_c^{(2)}(0)$(0.1), $g_c^{(2)}(0)$(0.25), and $g_c^{(2)}(0)$(0.5) is the calculated degree of second-order coherence via photon number distribution when the average number of photons in a detection time window equals 0.1, 0.25, and 0.5, respectively.}\label{t1}
\end{table}

The reasons why most of the calculated degree of second-order coherence via photon distribution, $g_c^{(2)}(0)$, is  less than the measured one via the HBT interferometer, $g_m^{(2)}(0)$, are as follows. As mentioned before, the response time of the detection system should be much less than the coherence time of the light in order to measure the accurate value of the degree of second-order coherence of light \cite{mandel-po}. Otherwise, photons in different modes may have to be taken into account, which will cause the measured photon distribution different from the real one. In our experiments, the coherence time of superbunching pseudothermal light is measured to be 1.28 and  4.63 $\mu$s. We will take the shorter one for discussion. The response time of the employed single-photon detector (SPCM-AQRH-14-FC, Excelitas Technologies) is 0.35 ns, which is much shorter than 1.28 $\mu$s, even if the 35 ns dead time between two single photon detections events was taken into account. The time resolution of the employed two-photon coincidence count detection system (DPC230, Becker \& Hickl) is 165 ps. Hence the response time of the whole system in the HBT interferometer is much less than the coherence time of the measured superbunching pseudothermal light. The measured value of degree of second-order coherence, $g_m^{(2)}(0)$, can be approximately treated as the accurate value of the degree of second-order coherence. On the other hand, the time window in the measurement of photon distribution is chosen to be 5  $\mu$s, which is longer than the coherence time of superbunching pseudothermal light. Hence the calculated degree of second-order coherence, $g_c^{(2)}(0)$, may be less than the accurate value due to the long collection time has to be taken into account \cite{mandel-po}. However, the average number of photons per measurement time window is less than 1. Photons from different modes contributing to the same detection time window may have less affect on the measurement results. This conclusion is supported by the  consistency between the value of $g_m^{(2)}(0)$ and $g_c^{(2)}(0)$ showing in Table \ref{t1}.

The reason why superbunching pseudothermal light generates Non-Rayleigh temporal speckles is the superposed larger number of field are correlated, which is similar as the way of generating Non-Rayleigh spatial speckles \cite{cao-2014, cao-2017, cao-2018, zhang-2019}. The intensity of laser light on the rotating groundglass is always changing in time, which makes the generated speckles are non-Rayleigh in the temporal domain. On the other hand, there is no correlation between the intensities at every instant between any two scatters on the rotating groundglass. The generated speckles are Rayleigh speckles in the spatial domain, even though the intensities of the scattered light at every point changes in time. Hence superbunching pseudothermal light can be employed to generate non-Rayleigh temporal speckles.

\section{Conclusions}\label{conclusions}

In conclusion, we have measured the photon distribution of superbunching pseudothermal light and found that the measured photon distribution is different from the one of pseudothermal light. Compared to the geometrical distribution with the same mean photon numbers, the photon number distribution is more distributed in the tail part. The larger the value of the degree of second-order coherence of superbunching pseudothermal light is, the higher the probability of detecting more photons in one measurement time window will be. We also calculated the degree of second-order coherence of pseudothermal and superbunching pseudothermal light via the measured photon distributions. It is found that the calculated values are consistent with the direct measured value of degree of second-oder coherence via a HBT interferometer. The results are helpful to understand the physics of two-photon bunching and superbunching. It is also suggested that superbunching pseudothermal light may be another method to generate Non-Rayleigh temporal speckles.

\section*{Acknowledgments}
This project is supported by Shanxi Key Research and Development Project (Grant No. 2019ZDLGY09-08), Open fund of MOE Key Laboratory of Weak-Light Nonlinear Photonics (OS19-2), and the Fundamental Research Funds for the Central Universities.

%%%%%%%%%% If using BibTeX:
\bibliography{sample}

%%%%%%%%%% If preparing manually:

\end{document}